\documentclass[prl,aps,twocolumn,showpacs]{revtex4}

\usepackage{graphicx}
\usepackage{dcolumn}
\usepackage{bm}

\begin{document}

\title{Crossovers in spin-boson and central spin models}

\author{P. C. E. Stamp$^{1}$, I. S. Tupitsyn$^{2}$}

\affiliation{ $^{1}$ Canadian Institute for Advanced Research,\\
and Physics Dept., University of British Columbia, 6224
Agricultural Rd., Vancouver, B.C., Canada V6T 1Z1\\
$^{2}$  Russian Science Centre "Kurchatov Institute", Moscow
123182, Russia }


\begin{abstract}

We discuss how the crossovers in models like spin-boson model are
changed by adding the coupling of the central spin to localised
modes- the latter modelled as a 'spin bath'. These modes contain
most of the environmental entropy and energy at low $T$ in
solid-state systems. We find that the low $T$ crossover between
oscillator bath and spin bath dominated decoherence, occurring as
one reduces the energy scale of the central spin, is characterised
by very low decoherence- we show how this works out in practise in
magnetic insulators. We then reconsider the standard
quantum-classical crossover in the dynamics of a tunneling system,
including both spin and oscillator baths. It is found that the
general effect of the spin bath is to broaden the crossover in
temperature between the quantum and classical activated regimes.
The example of tunneling nanomagnets is used to illustrate this.

\end{abstract}

\maketitle

\vspace{1cm}




\section{1: Introduction}
\label{sec:in}


In the book of Weiss on quantum dissipative phenomena
\cite{weiss99} one finds a very nice summary of results on the
crossover between quantum tunneling and classical activation for a
single tunneling coordinate coupled to a bath of oscillators (see
Chapters 10-17, particularly 14 and 16) . This kind of problem has
a long and interesting history, beginning with work of Kramers
\cite{kramers40} in 1940. The oscillator bath models assume that
each bath mode is weakly perturbed, and then the description of
the bath by oscillators is well known to be correct. Many physical
systems are very accurately described by such models
\cite{weiss99,hanggi90,kagAJL92,ajl87}, and they are central to
much of reaction rate chemistry as well. Typically one studies
either a particle tunneling from a trapped state to an open
continuum of states (the dissipative tunneling problem), or a
double-well system in which a particle has to go from one well to
another (the dissipative 2-well problem). One has a range of
temperatures in which both activation and tunneling processes are
important. Both the width of the crossover regime and the detailed
dependence of transition rates, as a function of temperature and
applied bias, are of interest \cite{weiss99,schmid92}. In the 2
well problem, the 'quantum limit', where only the 2 lowest levels
of the 2-well system are relevant (assuming a weak bias between
the wells), has been studied very extensively. This is the
'spin-boson model', in which a 2-level system couples to the
oscillator environment.

Another interesting application of the spin-boson model is to the
problem of qubits in quantum information processing (QUIP). The
central issue here is the study of decoherence in the dynamics of
the qubit, and how it depends on both simple things like applied
fields, temperature, etc., and in a more complex way on the
detailed nature of the bath, and its coupling to the qubit.  It
turns out that at the low temperatures that are appropriate for
QUIP, or for any other large scale quantum coherence, the
oscillator bath models are no longer adequate to describe all the
physics. In many systems the decoherence is controlled largely by
the coupling to localised modes, such as defects, tunneling
charges, paramagnetic spins, or nuclear spins, and this
environment of localised modes cannot in general be modelled by
oscillators- it can however be described as a set of spins
\cite{PS00} (the 'spin bath'). There is now extensive experimental
evidence for the key role of such modes in experiments on Cooper
pair box qubits \cite{naka02}, SQUID qubits
\cite{vdWal00,vdWalTh01}, and in molecular magnets
\cite{WWNS,morello03}, but the importance of these modes is
already rather obvious just from an estimation of their coupling
to these systems. There have been a fair number of theoretical
studies of spin bath environments. Early partial studies, in
various contexts, include refs.
\cite{shen67,davydov72,sta88,shimshoni92}; later work has
concentrated on application to coherence and relaxation in
tunneling systems (see, eg., refs.
\cite{PS00,PS93,PS96,castro93,dauriac96,levine97}), to decoherence
in mesoscopic conductors \cite{imry,vDelft}, superconducting
qubits \cite{PS00,tian99}, and nanomagnets \cite{tupPRL03}. It is
clear that in the limit of weak coupling to the spin bath, it
should be possible to map to an oscillator bath- studies in this
limit appear in, eg., refs. \cite{castro93,shao98,makri99,PS00}.

One can think of this breakdown of the oscillator bath model in
several ways. One is dealing here with a breakdown of the
assumption of weak system-bath couplings, and a corresponding
breakdown of linear response in the behaviour of the bath dynamics
(for more detailed discussion of this see ref. \cite{PS00}, and
refs. therein, and also the nice short summary by Weiss, in ref.
\cite{weiss99}, pp. 49-52). We emphasize that the coupling to
localised bath modes is almost always weak compared to the
tunneling barrier energy $E_B$, or to the energy $\omega_o$
corresponding to small oscillation energies in the potential
wells- so these modes are usually invisible in ordinary tunneling
experiments. However, the energy scale of the localised bath
modes, and their coupling to the central qubit coordinate, is
often {\it not} small compared to the exponentially smaller
tunneling energy $\Delta_o$. In the qubit regime it is the
comparison with $\Delta_o$ that counts, particularly for
decoherence (for a more precise discussion see below).

We are thus left with an interesting problem. What is the combined
effect of  spin bath and oscillator bath modes on the dynamics of
the central system? In particular, how is decoherence affected by
these two, and how does the oscillator bath take over from the
spin bath as one goes to higher temperatures, or increases
$\Delta_o$? Essentially one has to reconsider the whole question
of the crossover between quantum and classical regimes when both
baths are included.

In this paper we give a progress report on these questions for 2
kinds of crossover, viz:

(i) The crossover between spin bath controlled decoherence, which
dominates when $\Delta_o$ is small, and the oscillator
bath-controlled decoherence, which dominates when $\Delta_o$ is
large (in both cases, assuming low temperature). The most
interesting behaviour is in the crossover regime itself, when the
decoherence goes to a {\it minimum}. Thus by raising $\Delta_o$
one can go from an incoherent regime, through a regime of coherent
qubit dynamics, and then back to incoherent tunneling. To
illustrate the idea we show how the general idea works for
nanomagnets coupled to nuclear spins and phonons- The detailed
application to specific magnetic and superconducting systems is
discussed elsewhere \cite{tupPRL03,sqL}.

(ii) We look at how the spin bath influences the crossover between
the quantum tunneling and classical thermally activated regime.
This also involves a crossover between spin bath and oscillator
bath environments. Given the complexity of this crossover, we do
not attempt any complete discussion, but instead make some
qualitative remarks on the physics, and then present some results
for magnetic insulators (again involving phonons and nuclear
spins). For related work one may go to a series of papers
\cite{tupCross} on the application to ensembles of tunneling
magnetic molecules.


\section{2: Crossovers from Coherence to Incoherence}
\label{sec:2}


The spin-boson model has a control parameter $\Delta_o$ (the
operating frequency of the qubit); and we consider here the
crossover between the small $\Delta_o$ regime, where decoherence
is controlled by the spin bath, and the large $\Delta_o$ regime,
where it is controlled by the oscillator bath. The interesting
thing is that in the crossover between these 2 regimes lies a
'dead zone' where decoherence can be very low. This 'coherence
window' will be very important for solid-state based quantum
information processing.

\subsection{2a: QUBIT COUPLED TO OSCILLATOR AND SPIN BATHS}
\label{sec:2a}

We consider a 2-level system (a qubit) with the usual bare
Hamiltonian
\begin{equation}
{\cal H}_o = \Delta_o \hat{\tau}_x + \epsilon_o \hat{\tau}_z
 \label{Ho}
\end{equation}
This is coupled to both spin and oscillator baths. The thermal
energy $k_BT$, and the longitudinal and transverse field energies
$\epsilon_o, \Delta_o$, are assumed to be much less than the
energy gap $E_g$ to any higher levels of the system. In a magnetic
qubit (eg., a magnetic molecule, or a rate earth ion), this 'spin
gap' is typically $5-10~K$, and in a superconducting qubit the
corresponding Josephson plasma frequency depends strongly on the
junction geometry, and might be a little less.

The baths themselves are assumed to have Hamiltonians
\cite{weiss99,ajl87,PS00}:
\begin{eqnarray}
{\cal H}_o^{osc} &=& {1 \over 2} \sum_q \left[{p_q^2 \over m_q} +
m_q \omega_q^2 x_q^2 \right] \\
{\cal H}_o^{SB} &=& \omega_k^{\perp} \hat{\it m}_k \cdot \sigma_k
\;+ \; \sum_{kk'} V_{kk'}^{\alpha \beta} \sigma_k^{\alpha}
\sigma_{k'}^{\beta}
 \label{Hbath}
\end{eqnarray}
in terms of a set of oscillators $\{ x_q \}$ describing
delocalised modes and a set of spins $\{ \sigma_k \}$ describing
localised modes (here for simplicity assumed to be a set of Pauli
spin-$1/2$ systems). We have written the set of 'fields' $\{ {\bf
h}_k \}$, acting on the individual bath spins, in the form ${\bf
h}_k = \omega_k^{\perp} \hat{\it m}_k$, where  $\hat{\it m}_k$ is
a unit vector in the direction of the field.  We assume that a UV
cutoff $\Omega_o$ exists in these Hamiltonians, so that all spin
and oscillator degrees of freedom have energy $< \Omega_o$. The 2
baths are coupled to the central qubit via the following diagonal
couplings:
\begin{equation}
{\cal H}_{int} = \hat{\tau}_z \left[ \sum_q c_q^{\parallel} x_q \;
+ \; \sum_k \omega_k^{\parallel} \hat{\it l}_k \cdot
\hat{\sigma}_k \right]
 \label{Hint}
\end{equation}
where $\{ \hat{\it l}_k \}$ are a set of unit vectors. There can
also be non-diagonal couplings, ie., terms which operate only when
the qubit is switching between the eigenstates $\vert \uparrow
\rangle$ and $\vert \downarrow\rangle$ of $\hat{\tau}_z$. These
are usually specified by modifying the form of the transverse term
$\Delta_o \hat{\tau}_x$ in the bare qubit Hamiltonian. For the
oscillator bath one adds a coupling \cite{kagNVP92}
\begin{equation}
{1 \over 2} (\hat{\tau}_+ c_q^{\perp} x_q + H.c.)
 \label{NDosc}
\end{equation}
and for the spin bath one makes the substitution
\begin{equation}
\Delta_o \hat{\tau}_x \; \rightarrow \; {1 \over 2} \left[
\Delta_o \hat{\tau}_+ e^{i \sum_k \alpha_k {\bf n}_k \cdot
\vec{\sigma}_k} + H.c. \right]
 \label{NDspin}
\end{equation}
where the $\{ {\bf n}_k \}$ are unit vectors.

We briefly note the important features of these interactions.
First, we recall that the usual longitudinal spin-boson couplings
$\{ c_q^{\parallel} \}$ are typically $\sim O(N_o^{-1/2})$, where
$N_o$ is the number of oscillator degrees of freedom in the
Hilbert space (defined by the UV cutoff $\Omega_o$). On the other
hand the spin bath couplings $\{ \omega_k^{\parallel} \}$ may have
a quite different dependence- in magnetic qubit systems they are
usually independent of $N_s$, the number of spins, whereas in
SQUID qubit systems one has $\omega_k^{\parallel} \sim
O(N_s^{-2})$, at least in the simplest designs. For large $N_o$
this means that the oscillator bath couplings are very weak-
justifying the initial model of linear weak couplings. In the case
of a SQUID qubit coupled to a spin bath one sees that it ought to
be possible to map the problem to a spin-boson model, and indeed
one can \cite{PS00,sqL}. However this is not an option for
magnetic qubits- not only are the individual hyperfine couplings
between the qubit and the nuclear spins independent of the number
$N_s$ of nuclear spins, they are also large- in many cases
$\omega_k^{\parallel}$ for a single nuclear spin can exceed
$\Delta_o$! In this case we must deal directly with the spin bath,
and give up any hope of mapping the problem to a spin-boson model.

A second remark concerns the non-diagonal couplings. In cases
where the diagonal couplings happen to be zero (which can happen
under unusual circumstances) the non-diagonal couplings are the
only remaining decoherence mechanism- this makes them very
interesting for studies of decoherence (a point which has also
been noted in recent discussions of superconducting qubits
\cite{ioffe99}). On the other hand when the diagonal couplings are
non-zero, they usually dominate over the non-diagonal ones, at
least when the qubit is modelling a tunneling solid-state system.
It then follows that both $c_q^{\perp}/c_q^{\parallel}$ and
$\alpha_k$ are small- in fact $c_q^{\perp}/c_q^{\parallel} \sim
O(\Delta_o/\Omega_o)$, and $\alpha_k \sim
O(\omega_k^{\parallel}/\Omega_o)$ (for more details on this see
refs. \cite{PS96,PS00}).

Finally, we note that the interactions between bath modes are
treated differently in the oscillator and spin bath cases. In the
oscillator bath case it is is usually argued that any weak
anharmonic interactions have little relevance to the dissipation
or decoherence caused by the bath- that information and energy are
quickly transported away from the qubit, and so we can drop all
reference to intra-bath interactions. In the case of the spin
bath, however, it is clearly incorrect to drop such interactions-
even though they are usually very small. This is because the spin
bath describes local modes, which are not weakly coupled to the
qubit- accordingly a large amount of energy and information can in
principle be dumped into each mode and the $V_{kk'}^{\alpha \beta}
\sigma_k^{\alpha} \sigma_{k'}^{\beta}$ interaction is the only way
this can be redistributed. Over long time scales non-linear
effects become inevitable, and the size of $V_{kk'}^{\alpha
\beta}$ becomes very important. We shall see how this works below.

\subsection{2b: DECOHERENCE RATES}
\label{sec:2b}

We define the decoherence dynamics for the qubit in a fairly
standard way \cite{weiss99}, by assuming an initial state $\vert
\uparrow\rangle$, and calculating the reduced density matrix as a
function of time thereafter, once the spin and oscillator baths
are integrated out. The general form of the result (assuming the
bias $\epsilon_o = 0$ for simplicity) is
\begin{eqnarray}
&& 2 \rho (t) =  \nonumber \\
&& \left(
\begin{tabular}{cc}
$ ( 1 +  e^{-\Gamma_1(t)} \cos ( 2 \Delta_o t ))$  &
$ i e^{-\Gamma_2(t)} \sin ( 2 \Delta_o t )$ \\ \\
$ - i e^{-\Gamma_2(t)} \sin ( 2 \Delta_o t ) $ & $
( 1 -  e^{-\Gamma_1(t)} \cos ( 2 \Delta_o t )) $
\end{tabular}
\right) \nonumber \\
\label{rho}
\end{eqnarray}

Here $\Gamma_{\mu} (t)$, with $\mu = 1,2$, may have a complicated
time dependence. If $\Gamma_{\mu} (t) \rightarrow \Gamma_{\mu} =
const.$, (so that the coherence decays exponentially in time) one
can write $1/\Gamma_{\mu} \equiv T_{\mu}$, following NMR
terminology. In this case we say that, {\it in this basis}, the
decay rate $1/T_2$ of the off-diagonal matrix elements is the
decoherence rate- it characterizes the rate at which interference
between $\vert \uparrow\rangle$ and $\vert \downarrow\rangle$
states is lost. In other cases one can usually derive a
characteristic timescale $\tau_{\phi}$ for the loss of coherence,
and this is called the decoherence time. One also defines a
dimensionless measure of the decoherence rate, given by
\begin{equation}
\gamma_{\phi} = 1/\tau_{\phi} \Delta_o
 \label{gamma}
\end{equation}
or its inverse, the 'decoherence quality factor', often defined as
$Q_{\phi} = \pi/\gamma_{\phi} = \pi \tau_{\phi} \Delta_o$. This
Q-factor tells us roughly the number of coherent oscillations of
the system before decoherence sets in.

\vspace{3mm}

Here we first quickly recall the known results for decoherence in
this kind of problem. The dimensionless decay rate $\gamma_{\phi}$
has the following contributions:

\vspace{3mm}

{\bf (i) Oscillator bath contributions}: The decoherence rates
here depend on the Caldeira-Leggett spectral density
\cite{weiss99}. For the cases we are interested in one has the
following results:

(a) {\it Phonon decoherence}: this is relevant when we deal with
spin-phonon coupling in magnetic insulators. Typically one
considers a spin ${\bf S}$ (representing a molecular spin or other
nanomagnet), which truncates at low temperature to a magnetic
qubit (when $k_BT \ll \Omega_o$, where $\Omega_o$ is the spin gap
to the higher electronic excitations). For decoherence the most
important coupling to ${\bf S}$ is the non-diagonal coupling $\sim
S \Omega_o \vert {\bf q} \vert$ to acoustic phonons having a
Debye energy $\theta_D$ (for a simple derivation of this see ref.
\cite{PS96}, and for a thorough discussion of spin-phonon
couplings see \cite{pol96,abraBl}). Standard spin-boson methods
\cite{weiss99,ajl87}, applied to this coupling, give a
contribution $\gamma_{\phi}^{ph}$ to $\gamma_{\phi}$ of
perturbative (ie., golden rule) form; when the applied bias is
zero one gets \cite{PS96}:
\begin{equation}
\gamma_{\phi}^{ph} = [(S \Omega_o \Delta_o)^2/\theta_D^4]
\coth(\Delta_o/k_BT)
 \label{Dph}
\end{equation}
which is very weak at low energies (ie., for $\Delta_o \ll
\theta_D$). Although no qubit behaviour has yet been seen in
magnetic systems, there are extensive experimental results for the
effect of phonons on the spin dynamics of magnetic molecules.

(b) {\it Electronic decoherence}: This comes in when we need to
analyse decoherence in SQUID flux qubits
\cite{vdWal00,vdWalTh01,mooij99,chior03} or Cooper pair box charge
qubits \cite{naka02,naka99,vion02,naka03}. For example, in flux
qubits tunneling between flux states $\pm \phi_m$, and with
charging energy $E_c$, one has a dimensionless coupling $\alpha =
( 16 \phi_m^2 \omega_0 /E_c ) Q^{-1}$ between SQUID flux and
electronic bath, parametrised in terms of the SQUID $Q$-factor. In
this case of Ohmic dissipation the decoherence rate is
\cite{weiss99}
\begin{equation}
\gamma_{\phi}^e(\Delta_o) = {\pi \over 2} \alpha
\coth(\Delta_o/k_BT)
 \label{gE}
\end{equation}
where again we assume the system is in resonance.

\vspace{3mm}

{\bf (ii) Spin bath contributions}: We write the contributions to
the spin bath-induced decoherence in terms of the couplings
introduced in (\ref{Hint}) and (\ref{NDspin}). It is useful to
also introduce another quantity $E_o$ which quantifies the total
effect on a single qubit level of the coupling to the bath spins-
we have
\begin{equation}
E_o^2 = \sum_k (\omega_k^{\parallel})^2
 \label{Eo}
\end{equation}
so that $E_o$ is just the half-width of the Gaussian envelope of
$2^{N_s}$ spin bath states associated with each qubit state. This
formula is easily generalised to include 'higher spin' bath spins
(see below). In section 3 we say more about the structure of these
bath spin states inside this $2^{N_s}$-fold manifold.

There are 3 spin bath contributions to the decoherence
\cite{PS96,PS00}:

(a) {\it Noise decoherence}: spin diffusion inside the spin bath
causes the longitudinal bias acting on $\hat{\tau}_z$ to fluctuate
over a range $\Gamma_M$ in energy bias space, causing phase noise.
There will be a characteristic timescale $T_2$ associated with
this noise (over longer timescales the bias fluctuates over a
larger range- see section 3 for more details). If $\Delta_o^3 \ll
T_2^{-1} \Gamma^2_M$, the noisy bias fluctuates rapidly compared
to the molecular tunneling dynamics, causing incoherent tunneling
(this is the 'fast diffusion' limit for the spin bath
\cite{PS00}). This case is illustrated in Fig. 1, which shows the
way the two qubit levels are affected in time by the fluctuating
bias. In the opposite extreme of large $\Delta_o$ one has a much
smaller noise contribution to $\gamma_{\phi}$ of
\begin{equation}
\gamma_{\phi}^N = N_{eff}/\pi \Delta_o T_2
 \label{Q_N}
\end{equation}
where $N_{eff}$ is the number of bath spins which are active (the
exact number depends on the particular system- typically $N_{eff}
\sim O(N)$, but the exact fraction $N_{eff}/N$ can vary widely
from one system to another). This noise contribution
$\gamma_{\phi}^N \ll 1$ (ie., it only weakly affects coherence).
The analogue of Fig. 1 for this case would show very small
fluctuations, which hardly affect the dynamics.

\begin{figure}[ht]
\centering
\vspace{-2.2cm}
\includegraphics[scale=0.40]{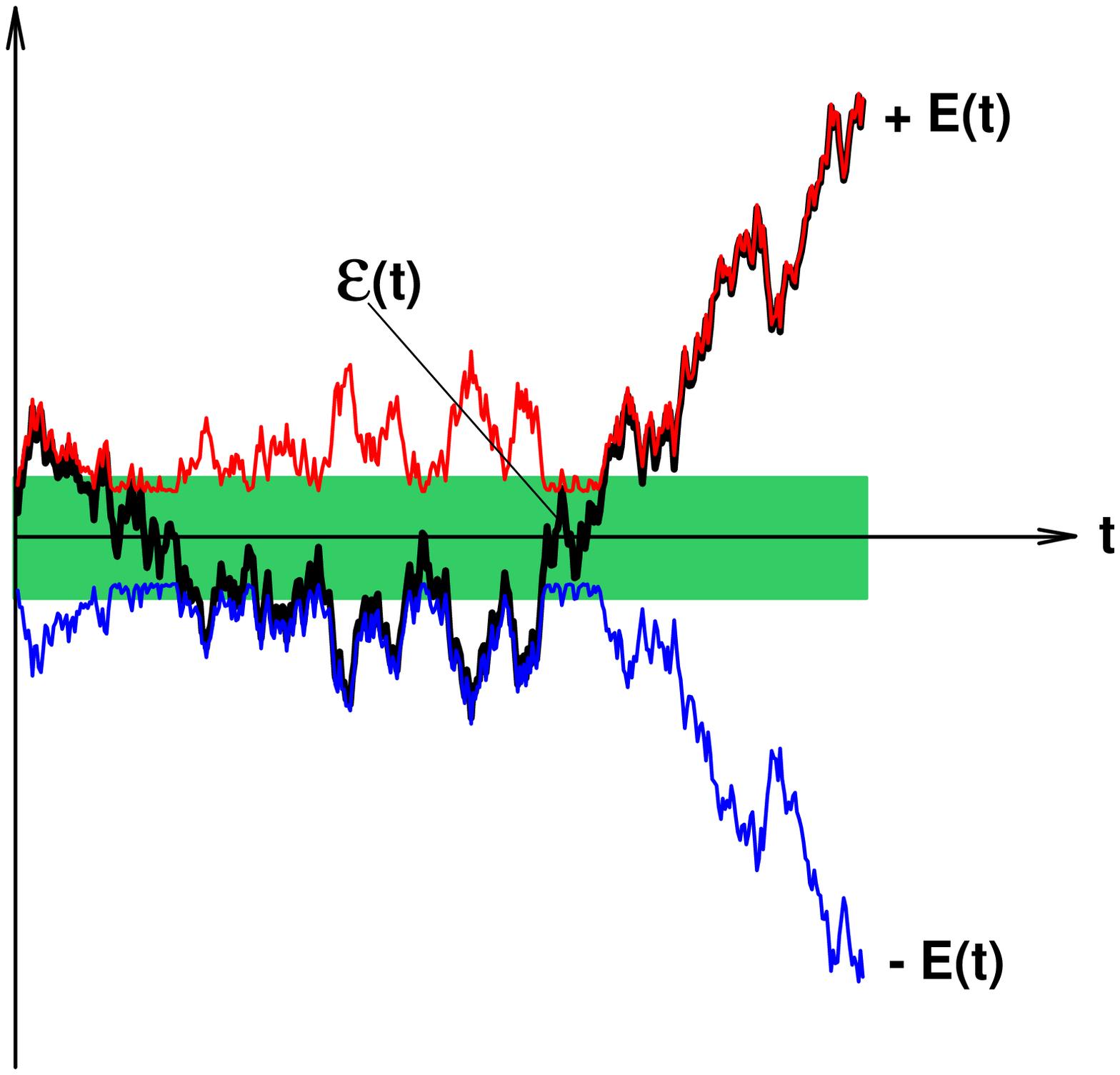}
\vspace{-2.8cm}
\caption{We show the effect of a randomly fluctuating
environmental noise bias $\varepsilon(t)$ (black curve) on a
tunneling 2-level qubit with tunneling matrix element $\Delta_o$.
The 2 levels having adiabatic energies $\pm E(t)$, with $E^2(t) =
\Delta_o^2 + \varepsilon^2(t)$, are shown as red \& blue curves.
The system can only make transitions when near "resonance" (ie.,
when $\vert \varepsilon(t) \vert$ is $\sim \Delta_o$ or less, the
regions shown in green). The resulting dynamics of the qubit is
incoherent in this case of strong noise.}
\label{fig:fig1}
\end{figure}

(b) {\it Precessional Decoherence}: The field about which the
$k$-th bath spin precesses changes each time the qubit flips, so
that the time evolution of the spin bath states becomes entangled
with that of the qubit. We can visualize this process by imagining
the precessional motion of a bath spin in the qubit field (Fig.
2). Integrating out the spin bath then gives decoherence in the
qubit dynamics. If the "operating frequency" $\Delta_o$ is low,
ie., $\Delta_o \ll E_o$, then this "precessional decoherence"
contribution $\gamma_{\phi}^{\kappa}$ to $\gamma_{\phi}$ is given
by
\begin{eqnarray}
\gamma_{\phi}^{\kappa} = {1 \over 2} \sum_k
(\omega_k^{\parallel}/\omega_k^{\perp})^2 \;\;\;\;\;\;\;\;\;\;\;\;
(if \;\; \omega_k^{\perp} \gg
\omega_k^{\parallel}, \Delta_o \;\;) \nonumber \\
\gamma_{\phi}^{\kappa} = {1 \over 2} \sum_k
(\omega_k^{\perp}/\omega_k^{\parallel})^2 \;\;\;\;\;\;\;\;\;\;\;\;
(if \;\; \omega_k^{\parallel} \gg \omega_k^{\perp}, \Delta_o
\;\;)\;
 \label{kapp1}
\end{eqnarray}
One gets the second result from the first by a duality, switching
the roles of $\omega_k^{\parallel}$ and $\omega_k^{\perp}$ in the
derivation of the first (cf. ref. \cite{PS00}, App. 2B).

If instead $\Delta_o \gg E_o$, ie., high operating frequency, then
also $\Delta_o \gg \omega_k^{\parallel}, \omega_k^{\perp}$. The
solution of this weak coupling problem is \cite{PS00,castro93}:
\begin{equation}
\gamma_{\phi}^{\kappa} = (E_o/\Delta_o)^2/2
 \label{kapp2}
\end{equation}
and this result is clearly important for the regime of coherent
qubit dynamics.

(c) {\it Topological Decoherence}: When the qubit flips, it causes
a sudden time-dependent perturbation on the bath spins, described
by the non-diagonal term (\ref{NDspin}). This induces transitions
in the bath spin states, and a corresponding contribution to the
entanglement of the bath spins and qubit states. Formally this
entangles the topological Berry phase of the qubit \cite{PS93}
with that of the bath spins, in the same way as for precessional
decoherence; after averaging over bath states the resulting
contribution $\gamma_{\phi}^{\lambda}$ to $\gamma_{\phi}$ is
\begin{equation}
\gamma_{\phi}^{\lambda} = {1 \over 2} \sum_k \vert \vec{\alpha}_k
\vert^2
 \label{lambda}
\end{equation}
where $\vert \vec{\alpha}_k \vert = \pi \vert \omega_k^{\parallel}
\vert /2 \Omega_o$ is assumed to be small (for general coupling
see refs. \cite{PS93,PS00}). In general this contribution is
smaller than the precessional decoherence.

\begin{figure}[ht]
\centering
\vspace{-0.8cm}
\includegraphics[scale=0.30]{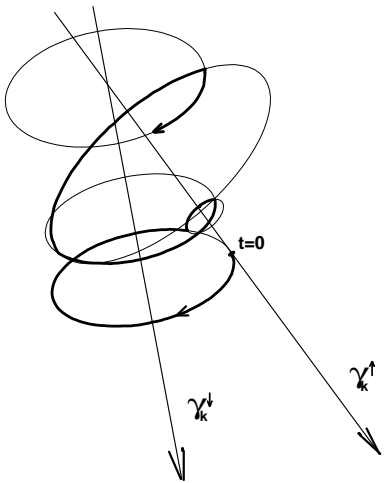}
\vspace{-1.2cm}
\caption{We show schematically the motion of a satellite
spin, in the presence of a qubit flipping between 2 different
states $\vert \uparrow \rangle$ and $\vert \downarrow \rangle$.
When the qubit flips, the qubit field acting on the $k$-th
satellite spin rapidly changes, from $\gamma_k^{\uparrow}$ to
$\gamma_k^{\downarrow}$ (or vice-versa). Between flips the spin
precesses around the qubit field, accumulating an extra
"precessional" phase. Averaging over this phase gives precessional
decoherence. The sudden change of qubit field also perturbs the
satellite spin phase, giving further decoherence (the "topological
decoherence" mechanism \cite{PS93}).}
\label{fig:fig2}
\end{figure}


We now observe, as has been noted before \cite{dube01}, that the
spin bath decoherence rate is always higher at low energy (small
$\Delta_o$), whereas the oscillator bath decoherence rate is
higher at high energy (large $\Delta_o$). Thus there will be a
'coherence window' at intermediate values of $\Delta_o$, where
$\gamma_{\phi}$ is small.

\subsection{2c: DECOHERENCE CROSSOVERS IN A MAGNETIC QUBIT}
\label{sec:2c}

At low $T$ the spin Hamiltonian of many large-spin nanomagnetic
systems (magnetic molecules, rare earth ions, or nanomagnetic
particles) reduces from that of a tunneling spin ${\bf S}$ to a
simple 2-state form ${\cal H}_o(\hat{\tau}) = (\Delta_o
\hat{\tau}_x + \epsilon_o \hat{\tau}_z)$, with the Pauli spin
$\hat{\tau}$ acting on the 2 lowest spin levels
\cite{wwRev,tupBB}, as in our qubit Hamiltonian (\ref{Ho}) . The
spin gap $E_g$ to the next levels is typically $\sim 5-10~K$, and
the 2-state picture is valid at energies $\ll E_g$. We assume
henceforth an "easy $\hat{z}$-axis" nanomagnet; then the 'bias'
energy $\epsilon_o = g \mu_B S_z H_o^z$. When $\epsilon_o = 0$, the
splitting $\Delta_o$ between the 2 "qubit" states $\vert -
\rangle, \vert + \rangle$ (bonding and anti-bonding eigenstates of
${\cal H}_o(\hat{\tau})$) is produced by tunneling between 2
potential wells, with each well having a "small oscillation"
energy $\Omega_o$; typically $\Omega_o \sim E_g$. The qubit is
thus the result of truncating out the higher spin states of the
nanomagnetic system, which we should schematically in Fig. 3.

\begin{figure}[ht]
\centering
\vspace{-2.8cm}
\includegraphics[scale=0.40]{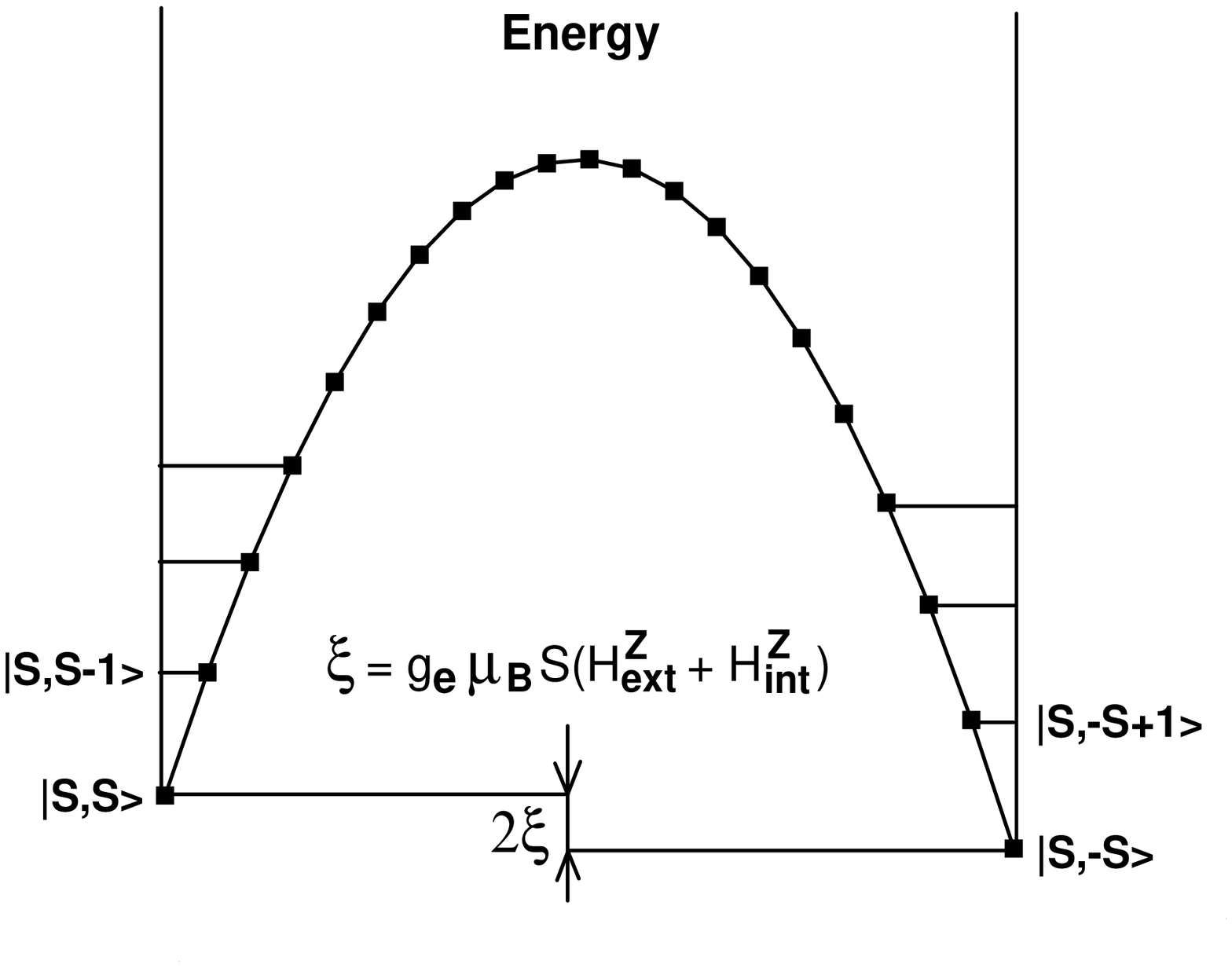}
\vspace{-3.6cm}
\caption{Magnetic anisotropy barrier of a small magnetic
system, such as a magnetic molecule. We show the eigenstates
$\vert S,m \rangle$ of the longitudinal part ${\cal H}_o^z(S_z)$
of the spin Hamiltonian ${\cal H}_o({\bf S})$. An external
longitudinal field $H^z_{ext}$ (or an internal field $H^z_{int}$),
biases the effective potential by an amount $\xi$. Adding
transverse anisotropy terms to the spin Hamiltonian causes
tunneling between the states in the figure. The 'ground state'
tunneling amplitude $\Delta_S$ between states $\vert S,S \rangle$
and $\vert S,-S \rangle$ is called $\Delta_o$ in the text.}
\label{fig:fig3}
\end{figure}

We define the states $\vert \uparrow \rangle,\vert \downarrow
\rangle$ (eigenstates of $\hat{\tau}_z$) by $\vert \pm \rangle =
\vert \uparrow \rangle \pm \vert \downarrow \rangle$. If the total
nanomagnetic spin ${\bf S}$ is not too small, these states
correspond roughly to semiclassical spin coherent states
\cite{auerbach94}, having orientations ${\bf n}_{\sigma}$ (here
$\sigma = {\uparrow},{\downarrow}$), which depend on both the
internal anisotropy field of the nanomagnet, and any transverse
external field ${\bf H}_o^{\perp}$. The splitting $\Delta_o$
depends sensitively on ${\bf H}_o^{\perp}$.

The intrinsic decoherence in insulating nanomagnets comes from
entanglement of the nanomagnetic spin wave function with that of
the nuclear spins and phonons \cite{PS96}. We first see how to
write these couplings in the form given in section 2(a). The
details for the spin bath are a slight generalisation of this
form, because the nuclear spins are not necessarily spin-$1/2$
objects.

The nuclear spins $\{ {\bf I}_k \}$ couple to the electronic spins
$\{ {\bf s}_j \}$ in ${\bf S}$ (where ${\bf S} = \sum_j {\bf
s}_j$) via hyperfine couplings
\begin{equation}
{\cal H}_{hyp} = A_{\alpha \beta}^{jk} s_j^{\alpha} I_k^{\beta}
 \label{hyp}
\end{equation}
whose form we do not specify here. We then define the
field-dependent quantity
\begin{equation}
\omega_k^{\parallel} = {1 \over 2I_k} \vert \sum_j A_{\alpha
\beta}^{jk} (\langle s_j^{\alpha} \rangle^{\uparrow} - \langle
s_j^{\alpha} \rangle^{\downarrow}) I_k^{\beta}\vert
 \label{w-par}
\end{equation}
where $\langle s_j^{\alpha} \rangle^{\sigma}$ is the expectation
value of ${\bf s}_j$ when ${\bf S} \rightarrow S{\bf n}^{\sigma}$.
The energy change of ${\bf I}_k$ when ${\bf S}$ flips from $S{\bf
n}_{\uparrow}$ to $S{\bf n}_{\downarrow}$ is then
$2I_k\omega_k^{\parallel}$, ie., there is a diagonal coupling
$\hat{\tau}_z \omega_k^{\parallel} \hat{\it l}_k \cdot \hat{\bf
I}_k$ between the qubit and ${\bf I}_k$, where $\hat{\it l}_k$ is
a unit vector parallel to the hyperfine field on ${\bf I}_k$. This
is just the coupling specified in (\ref{Hint}) in the last
section.

The external transverse field ${\bf H}_o^{\perp}$ couples to ${\bf
I}_k$ with Zeeman coupling $\omega_k \hat{\it m}_k \cdot \hat{\bf
I}_k$, where $\omega_k \hat{\it m}_k = g_k^N \mu_N {\bf
H}_o^{\perp}$ and $\hat{\it m}_k$ is a unit vector along ${\bf
H}_o^{\perp}$. This is the same as the coupling given in
(\ref{Hbath}). Finally, the interactions between the spins in the
spin bath add a term
\begin{equation}
{\cal H}_{NN} (\{ {\bf I}_k \}) = \sum_{k=1}^N \sum_{k'=1}^N
V^{\alpha \beta}_{kk'} \hat{I}^{\alpha}_k \hat{I}^{\beta}_{k'}
 \label{V-NN}
\end{equation}
whose effect on the qubit will be handled by assuming that the
spin diffusion caused by this weak interaction adds a "noise" term
$\xi(t) \hat{\tau}_z$ to the static bias $\epsilon_o
\hat{\tau}_z$. These terms taken together define an effective
interaction Hamiltonian ${\cal H}_{NS} = {\cal H}_o(\hat{\tau}) +
V(\hat{\tau}, {\bf I}_k)$, where
\begin{equation}
V =  \hat{\tau}_z \left[\xi^z(t) + \sum_k \omega_k^{\parallel}
\hat{\it l}_k \cdot \hat{\bf I}_k \right] + \sum_k \omega_k
\hat{\it m}_k \cdot \hat{\bf I}_k
 \label{Ham}
\end{equation}

To (\ref{Ham}) we also add a 'spin-boson' coupling
\begin{equation}
{\cal H}_{s \phi} = \sum_q c_q^{\perp}\hat{\tau}_x x_q \;\;\sim
\;\; \sum_q S \Omega_o |{\bf q}| x_q \hat{\tau}_x
 \label{Hsphi}
\end{equation}
to the acoustic phonon coordinate $x_q$. This non-diagonal term
was already discussed above.

The energy scale over which the nuclear spin bath states operate
is just the linewidth $E_o$ of the entire multiplet of nuclear
states coupling to each qubit level \cite{PS96}. It is clear that
if the energy bias $\xi$ in the problem is less than $E_o$, it
will be possible for the system to make tunneling transitions
without the aid of the phonons, even if $\xi \gg \Delta_o$ (see
Fig. 4). For this problem, with higher spin nuclei, one easily
finds that
\begin{equation}
E_o^2 = \sum_k {I_k + 1 \over 3I_k} (\omega_k^{\parallel}I_k)^2
 \label{EoN}
\end{equation}
The acoustic phonon energy scale is the Debye energy $\theta_D$.
Now in a nanomagnetic system the ratio $E_o/k_B \theta_D$ can be
$\lesssim 10^{-4}$, suggesting the a very simple tactic for
suppressing decoherence. If we tune $\Delta_o$ so that $k_B
\theta_D \gg \Delta_o \gg E_o$, then we will be in the "coherence
window" mentioned above, where the qubit dynamics is too slow to
disturb most phonons appreciably, but too fast for the nuclear
spins to react.

\begin{figure}[ht]
\centering
\vspace{-2.2cm}
\includegraphics[scale=0.35]{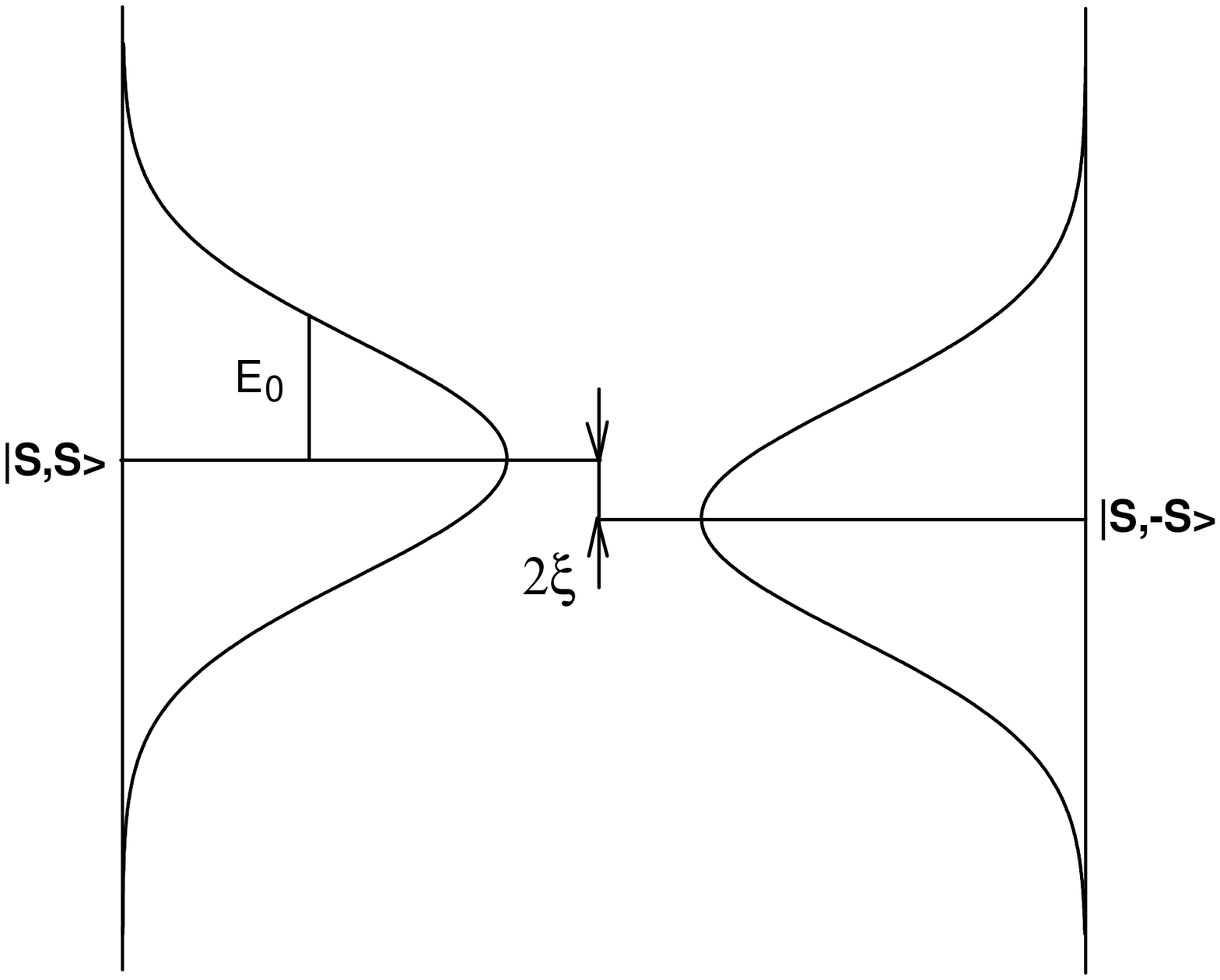}
\vspace{-2.6cm}
\caption{The 2 qubit levels connected with the zero-field
states $\vert S,S \rangle$ and $\vert S,-S \rangle$ each couple,
via the internal hyperfine interactions, to a a very large number
of nuclear spin levels. The result is a multiplet associated with
each qubit level, which usually has a Gaussian density of levels,
with a half-width $E_o$ (see text).}
\label{fig:fig4}
\end{figure}

To substantiate this idea, we generalise the low field ($\Delta_o
< E_o$) calculations of nanomagnetic dynamics \cite{PS96}, where
incoherent tunneling relaxation is found, to the high-field regime
$\Delta_o \gg E_o$. Because the $\{ \omega_k^{\parallel} \} \ll$
the Zeeman couplings $\{ \omega_k^{\perp} \}$, and $\{
\omega_k^{\parallel} \} \ll \Delta_o$, this dynamics can be solved
\cite{PS00,castro93}, by expanding (\ref{Ham}) in
$\omega_k^{\parallel}/\Delta$. If we first ignore the noise term,
we get a new effective Hamiltonian
\begin{eqnarray}
H_{NS} &=& \left[ \Delta_o + \sum_{kk'} {\omega_k^{\parallel}
\omega_{k'}^{\parallel} \over 2 \Delta_o} (\hat{\it l}_k \cdot
{\bf I}_k) (\hat{\it l}_{k'} \cdot {\bf I}_{k'}) \right]
\hat{\tau}_x \nonumber \\  &+&  \sum_k \omega_k^{\perp} \hat{\it
m}_k \cdot {\bf I}_k + O((\omega_k^{\parallel})^4/\Delta_o^3)
 \label{12a}
\end{eqnarray}
We can easily generalise the derivation of the result given in
(\ref{kapp2}) for $\gamma_{\phi}^{\kappa}$ to the case where the
nuclear spins have a spin modulus $I_k > 1/2$. One finds easily
that
\begin{eqnarray}
\gamma_{\phi}^{\kappa} &=& \sum_{kk'} \sqrt{{(I_k + 1)(I_{k'} + 1)
\over 9I_k I_{k'}}} {\omega_k^{\parallel} \omega_{k'}^{\parallel}
I_k I_{k'} \over 2 \Delta_o^2}  \nonumber \\
&=& {1 \over 2} \left( {E_o \over \Delta_o} \right )^2
\label{kapp2a}
\end{eqnarray}
Thus we see that the result (\ref{kapp2}) for precessional
decoherence is generally valid, regardless of the size of the
nuclear spins.

We now look at the 2 other contributions to the decoherence rate.
The nuclear spin transitions induced directly by electronic spin
flips add a contribution $\gamma_{\phi}^{\lambda}$ to
$\gamma_{\phi}$. However when $\Delta_o \gg E_o$, the ratio
$\gamma_{\phi}^{\lambda}/ \gamma_{\phi}^{\kappa} \sim
O(\Delta_o^2/\Omega_o^2) \ll 1$, ie., the precessional decoherence
always dominates over the topological decoherence.

The third contribution $\gamma_{\phi}^N$ comes from nuclear spin
noise (the term $\xi(t) \hat{\tau}_z$). When $\Delta_o \gg E_o$,
these fluctuations are extremely slow compared to $\Delta_o$;
typically $T_2 \sim msecs$ at low $T$, where $T_2$ is the typical
transverse nuclear relaxation time for the $N$ nuclei controlling
these fluctuations. One then gets $\gamma_{\phi}^N = N/\pi
\Delta_o T_2$; we will see this is very small.

This summarizes the nuclear spin terms. The phonon contribution is
as described above in (\ref{Dph}), and in the low $T$ limit we are
interested in, where $k_BT < \Delta_o$, we have:
\begin{equation}
\gamma_{\phi}^{ph} \;\; \rightarrow \;\; [(S \Omega_o
\Delta_o)^2/\Theta_D^4]
\label{Dph1}
\end{equation}
At temperatures above $\Delta_o$ the phonon decoherence rate
increases.

Now, since $\gamma_{\phi}^{\kappa}$ dominates nuclear spin
decoherence, we can get an estimate for the optimal decoherence
rate $\gamma_{\phi}^{min}$ by simply minimizing
$\gamma_{\phi}^{\kappa} + \gamma_{\phi}^{ph}$ with respect to
$\Delta_o$, assuming $k_BT < \Delta_o$, to get:
\begin{equation}
\gamma_{\phi}^{min} \approx \sqrt{2} S \Omega_o E_o / \theta_D^{2}
 \label{g-min}
\end{equation}
at an optimal tunneling splitting $\Delta_o^{opt}$:
\begin{equation}
\Delta_o^{(opt)} \approx \theta_D (E_o / \sqrt{2} S \Omega_o)^{1/2}.
 \label{D-opt}
\end{equation}
We see that decoherence is optimised for a given $S$ by making
$E_o$ and $\Omega_o$ small, and $\theta_D$ large, within the
constraint that $\Omega_o \gg \Delta_o > k_BT$. If $k_BT >
\Delta_o$ we get a different (less favorable) answer.

The detailed application of this kind of result to a nanomagnet,
in cases where one knows something about the couplings, is in
principle very useful for designing magnetic qubits. The tunneling
splitting is most easily modified just by applying a magnetic
field transverse to the easy axis- this can be used to tune
$\Delta_o$ over many orders of magnitude (see Fig. 5). For more
details of such applications, see ref. \cite{tupPRL03}.

\begin{figure}[ht]
\centering
\vspace{-2.0cm}
\includegraphics[scale=0.35]{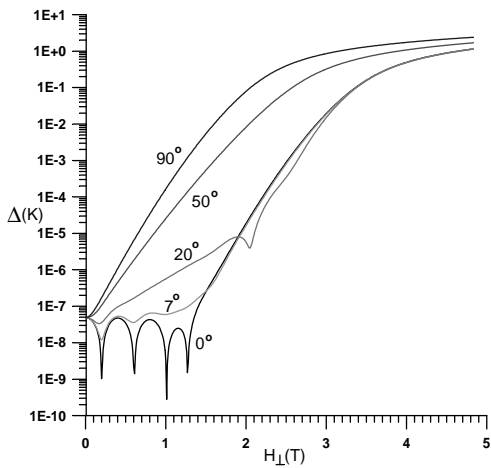}
\vspace{-1.6cm}
\caption{The tunneling splitting $| \Delta_o |$ in the $Fe$-8
molecule, which is a good example of a tunneling nanomagnet. The
tunneling anisotropy potential for the spin is biaxial, with an
easy $z$-axis and a hard $x$-axis. The tunneling splitting is
shown as a function of a transverse field $H_{\perp}$ oriented in
the $xy$-plane, at an angle $\phi$ from the hard axis.}
\label{fig:fig5}
\end{figure}

\section{3: Interlude-Spin Bath dynamics}
\label{sec:3}

In the results given above for decoherence rates, the intrinsic
dynamics of the oscillator and spin baths played only a secondary
role. In both cases it was assumed that phase information
exchanged between system and bath was lost once it was taken up by
the bath. This may not always be realistic, particularly in the
spin bath- one can easily imagine situations in which the spin
bath is cycled so as to recover some of this phase information,
and in NMR this is actually done (eg., in 'multiple quantum
coherence' experiments \cite{schlichter}). Even for oscillator
baths it is well known that anharmonic oscillator couplings can
allow the bath to hold information in certain modes for very long
times- now a very well-studied phenomenon \cite{fermi}.

The question of the intrinsic bath dynamics is also important when
one looks at the crossover between quantum and classical
relaxation (next section). Therefore here we clarify what is and
is not contained in the models we use.

As described by equation (\ref{Hbath}), the oscillator bath has
the very simple dynamics of $N$ independent oscillators, with
frequencies $\{ \omega_q \}$. Coupling to the 'central system'
hardly changes this dynamics, since the coupling strengths $c_q
\sim O(N^{-1/2})$. Thus nowhere in this model does energy and
phase relaxation occurs in the bath- both simply accumulate
independently in each mode. In reality anharmonic couplings cause
rapid relaxation in systems of extended modes like electrons,
phonons, magnons, etc; the only system to which (\ref{Hbath})
strictly applies is a bath of photons in a vacuum, in which the
very weak vacuum polarisation-induced photon-photon interactions
have been dropped. The reason that models like the spin-boson
model work in most (but not all!) cases is just because this
relaxation is usually fast- once energy or phase information has
gone from the central system into a particular mode, it is rapidly
diffused into other modes and so hard to recover.

In the case of the spin bath dynamics one has to be more careful.
As discussed in detail in ref. \cite{PS00}, one can classify the
large number of spin bath states into 'polarisation groups',
defined by their total polarisation $M = \sum_k \hat{z}_k \cdot
\sigma_k $ along some set of axes $\{ \hat{z}_k \}$, where the $\{
\hat{z}_k \}$ are unit vectors. If the 'external' field strengths
$\{ \omega_k^{\perp} \}$ are weak, so that $\omega_k^{\perp} \ll
\omega_k^{\parallel}$, then it makes sense to have these axes
along the direction of the local fields acting on the $\{ \sigma_k
\}$, coming from the qubit, ie., to make $\hat{z}_k = \hat{\it
l}_k$ (compare eqtn (\ref{Hint})). If on the other hand the
external fields dominate, one instead assumes that $\hat{z}_k =
\hat{\it m}_k$, ie., the axes of quantization defining $M$ are
just the external field directions. If these fields ${\bf h}_k =
\omega_k^{\perp} \hat{\it m}_k$ are indeed defined by some strong
external magnetic field ${\bf H}_o = \hat{\it n}_o H_o$, then
$\hat{z}_k = \hat{\it n}_o$ for all spins in the bath, ie., the
axis defining the bath polarisation is just the external field
direction. However we emphasize that in many cases the fields
${\bf h}_k$ may have nothing to do with any external magnetic
field. For example, the $\{ \sigma_k \}$ might refer to a set of
defects in a glass- in this case they couple to a strain field
like electric dipoles. Even in the case of real spins the ${\bf
h}_k$ may not be external fields but internal ones. For example,
many tunneling nanomagnets do not tunnel between oppositely spin
states, because the magnetic anisotropy field does not have
uniaxial symmetry. In this case the hyperfine field on the nuclear
spins does not flip through $180^o$ when the nanomagnet tunnels.
One then resolves the hyperfine field into 2 components; ${\bf
h}_k$ defines that component which does not change during the
flip, while $\omega_k^{\parallel} \hat{\it l}_k$ defines a
component which flips through $180^o$.

The point of defining polarisation groups in this way is that it
then makes sense to define 2 relaxation times $\tilde{T}_1$ and
$\tilde{T}_2$, such that $\tilde{T}_1$ defines the relaxation of
$M$, and $\tilde{T}_2$ the spin bath relaxation within a given
polarisation group. Note that even if the spin bath is actually
made up of nuclear spins, these times are ${\it not}$ the
$T_1,T_2$ times measured in a typical NMR experiment, which only
looks at a single nuclear species at a time- the times here refer
to the whole spin bath. One can imagine an NMR experiment which
polarizes a particular nuclear species, and then observes rapid
relaxation of this polarisation into other nuclear species, via
the inter-spain interactions- even though the total spin along the
field is conserved (so that the spin bath $\tilde{T}_1$ defined
here is still infinite).

Usually one expects $\tilde{T}_1 \gg \tilde{T}_2$ at low $T$,
because the interspin couplings $V_{kk'}^{\alpha \beta}
\sigma_k^{\alpha} \sigma_{k'}^{\beta}$ can mediate transverse
spin-spin relaxation processes, ie., cause spin diffusion in the
spin bath and contribute to $\tilde{T}_2$, whereas $\tilde{T}_1$
processes typically require interaction with some external system.
The most typical case where the bath spins inter-communicate by
magnetic or electric dipolar interactions is actually hard to
analyse theoretically, because these interactions are marginal in
3 dimensions (integrals of the form $\prod_j^{n-1} \int d^3 {\bf
r}_j/r_j^3$ appear in the calculation of the relaxation dynamics
of $n$ coupled bath spins, so that multi-spin couplings are just
as important as pairwise interactions, and distant spins as
important as nearby ones \cite{levitov90}). Thus we would not
usually try to calculate the spin bath $\tilde{T}_2$.

Let us now emphasize one of the crucial differences between the
spin and oscillator baths. This is that because the $\{
\omega_k^{\parallel} \}$ are not weak, the spin bath dynamics
depends very strongly on what the central system is doing. In the
case of a central qubit one can imagine 2 extreme scenarios:

(i) We freeze the qubit dynamics by applying a longitudinal field
$\epsilon_o \gg \Delta_o$. Then the spin bath dynamics is
described by the effective Hamiltonian
\begin{equation}
{\cal H}_{SB} = \sum_k {\bf b}_k \cdot \sigma_k \;+ \; \sum_{kk'}
V_{kk'}^{\alpha \beta} \sigma_k^{\alpha} \sigma_{k'}^{\beta}
 \label{H-SB}
\end{equation}
where the static fields $\{ \hat{\bf b}_k \}$ are given by
\begin{equation}
{\bf b}_k \;=\; \omega_k^{\perp} \hat{\it m}_k  \pm
\omega_k^{\parallel} \hat{\it l}_k
 \label{b_k}
\end{equation}
with the sign $\pm$ depending on whether the central qubit is
frozen in the $\vert \uparrow\rangle$ or $\vert \downarrow\rangle$
state.

We see that in this extreme case the bath would have its total
polarisation conserved, provided we defined the polarisation
groups using axes $\{ \hat{z}_k \}$ parallel to the $\{ \hat{\bf
b}_k \}$. Actually we would not normally do this, but one can
easily imagine a situation in which the applied field is either
very strong or very weak, and then $M$ would be almost exactly
conserved, ie., $\tilde{T}_1$ would be very long.

Now suppose we switch on the qubit dynamics- the easiest way to do
this is to remove the bias field $\epsilon_o$. Then the nuclear
bath finds itself subject to a quite different time-dependent
Hamiltonian, of the form
\begin{eqnarray}
{\cal H}_{SB}(t) &=& \sum_k \omega_k^{\perp} \hat{\it m}_k \cdot
\sigma_k \;+ \; \sum_{kk'} V_{kk'}^{\alpha \beta}
\sigma_k^{\alpha} \sigma_{k'}^{\beta} \nonumber \\
&+& \tau_z (t) \sum_k \omega_k^{\parallel} \hat{\it l}_k \cdot
\hat{\sigma}_k
 \label{b-ka}
\end{eqnarray}
where the time-dependent variable $\tau_z (t)$ is jumping back and
forth between $\pm 1$ with some correlation time $\sim
1/\Delta_o$. This Hamiltonian described the spin bath now subject
to an external 'telegraph noise', which causes transitions between
different polarisation groups of the system- this will be true no
matter how these polarisation groups are defined, provided both
the $\{ \omega_k^{\perp} \}$ and the $\{ \omega_k^{\parallel} \}$
are non-zero. Suppose, for example, that we have a weak external
field. Then $M$ is defined as $M = \sum_k \hat{\it m}_k \cdot
\sigma_k$, as discussed above. We can imagine an initial state
where all bath spins are oriented parallel or antiparallel to the
initial local fields- but as soon as the qubit flips, they begin
to precess. As discussed above, this is what causes precessional
decoherence.

However, as observed already above, the motion of the central
qubit depends itself on the spin bath dynamics- we must never
forget that the telegraph noise acting on the spin bath depends in
turn on the spin bath state. The simplifying feature is that
energy conservation imposes a simple constraint on the allowed
bath dynamics. In general the qubit will be off resonance by some
energy $\xi$, the sum of an external field contribution and the
internal field from the spin bath:
\begin{equation}
\xi = \epsilon_o + \sum_k \omega_k^{\parallel} {\it l}_k \cdot
\sigma_k
 \label{xi}
\end{equation}
In the simplest case where the couplings $\omega_k^{\parallel}$
are either dominated by a single value $\omega_o^{\parallel}$, or
else all cluster around this value (this happens in, eg., rare
earth magnets like the $LiHo_xY_{1-x}F_4$ system \cite{LiHo},
where the $Ho$ hyperfine coupling to the $Ho$ nuclear spin is much
larger than its coupling to the other nuclear spins), we have
approximately that $\xi \sim \epsilon_o + \omega_o^{\parallel} M$
(with some spread around the value $\omega_o^{\parallel}$, caused
by dispersion in the values of the $\omega_k^{\parallel}$). Now
for the qubit to make transitions we require that the initial and
final energies $\pm \xi$ (for $\tau_z = \pm$) be the same within
$\Delta_o$. This means that if $\epsilon_o \gg \Delta_o$, the
central qubit can only flip if the spin bath absorbs the extra
energy. However this will in general involve a change of $2M$ in
the net bath polarisation, such that $\omega_o^{\parallel}\vert M
\vert \sim \vert \epsilon_o \vert$, each time the system flips.
Thus at least $M$ bath spins have to flip- the time-varying field
of the qubit must drive these transitions.

A formal calculation of the spin bath dynamics incorporates this
constraint using a projection operator \cite{PS00}, involving a
dummy variable $\xi$:
\begin{equation}
{\hat \Pi}_M = \delta (\sum_{k=1}^N {\hat \sigma }_k^z -M) =
\int_0^{2\pi} {d\xi \over 2\pi} e^{ i\xi ( \sum_{k=1}^N {\hat
\sigma }_k^z -M) } \;.
 \label{b.4}
\end{equation}
which restricts all bath states to the $M$-th polarisation group.
Suppose now that the spin bath starts off with polarisation $M =
M_o$, and that to maintain resonance, the net polarisation must
change by $\pm 2M$ each time the qubit flips from
$|\uparrow\rangle$ to $|\downarrow\rangle$ or vice-versa, ie., it
cycles between $M_o \longleftrightarrow M_o - 2M$. The simplest
case arises if we ignore the interaction between bath spins
completely, ie., let $V_{kk'}^{\alpha \beta} \rightarrow 0$ in
(\ref{b-ka}). Then one can write the dynamics of the bath in terms
of operators $\hat{T}_{n}$ and $\hat{U}_k$, acting on the spin
bath in the presence of $n$ flips of the qubit. The $\hat{T}_n$
are given by
\begin{eqnarray}
 {\hat T}_{n} &=&\bigg[ e^{i\xi_{n}\sum_{k=1}^N {\hat \sigma }_k^z}
{\hat U}^{\dag } e^{i\xi_{n-1}\sum_{k=1}^N {\hat \sigma }_k^z}
{\hat U}   \nonumber \\ & ...& {\hat U}^{\dag } e^{i\xi_{1}
\sum_{k=1}^N {\hat \sigma }_k^z} {\hat U} \bigg] \;.
 \label{b.6}
\end{eqnarray}
(involving a set of $n$ dummy variables) and the $\{ \hat{U}_k \}$
define the change in the wave-function of the bath spins caused by
the sudden flip of the qubit from one orientation to another, ie.,
the mismatch between in and out states:
\begin{equation}
\mid \{ {\vec \sigma }_k^{out} \} \rangle = \prod_{k=1}^N {\hat
U}_k \mid \{ {\vec \sigma }_k^{in} \} \rangle
 \label{b.3}
\end{equation}
Suppose, eg., that $\omega_k^{\perp} \ll \omega_k^{\parallel}$,
for all bath spins, ie., the field on each bath spin almost
exactly reverses during each flip, through an angle $180^o -
2\beta_k$. Then $\hat{U}_k$ is just
\begin{equation}
{\hat U}_k =  e^{ -i\beta_k {\hat \sigma }_k^x }
 \label{Uk}
\end{equation}

Suppose we now want to write down the amplitude for the spin bath
to start in the polarisation group $M_o$ and finish in the same
polarisation group. This can only happen if the qubit flips $2n$
times. The amplitude is then the sum of a term
\begin{eqnarray}
G^{\uparrow\uparrow}_{M_o,M} (t) &=&   {(i\Delta_o t)^{2n} \over
(2n)!} \prod_{i=1}^{2n}  \int {d\xi_i \over 2\pi } e^{-i M_o
(\xi_{2n}+\xi_{2n-1}+\dots +\xi_1 )} \nonumber \\
& \times &  e^{2iM(\xi_{2n-1}+\xi_{2n-3}+\dots +\xi_1 )} {\hat
T}_{2n}
 \label{b.5}
\end{eqnarray}
acting on the initial state of the spin bath, in which the qubit
is assume to start and finish in the same state
$|\uparrow\rangle$, and another terms in which it starts and
finishes in the state $|\downarrow\rangle$.

The result of form (\ref{b.5}) is only complete if energy
conservation requires that the polarisation change by $\pm M$ each
time. However in most cases there will be a wide range of values
of $\omega_k^{\parallel}$, rather than a single dominant value,
and so each polarisation group will be widely spread in energy
space, and a large number of polarisation groups will have states
at a given energy $\xi$ (ie., the groups will strongly overlap in
energy space). In this more general case we should sum over
transition amongst these groups with the appropriate weighting-
the details are an obvious extension of what has just been
described. In this way we can give a theoretical evaluation of the
time $T_1$.

To calculate the full dynamics of the spin bath we must also
include the action of the interspin interaction $V_{kk'}^{\alpha
\beta}$. This enables transitions amongst the different bath
states inside the same polarisation group, even when $M$ does not
change, ie., an evaluation of the spin bath $T_2$. We do not
discuss here how such calculations may be done.

We underline here again the most important point of this
interlude- that the spin bath controls the qubit dynamics,
deciding whether the qubit may flip or not- but in its turn the
qubit drives the spin bath dynamics.


\section{4: Quantum to Classical crossover}
\label{sec:4}

As noted in the introduction, a great deal is known about how the
dynamics of a single quantum system, coupled to a thermal bath,
changes as one raises the bath temperature
\cite{weiss99,hanggi90}. Many analyses use an oscillator model to
describe the bath. However the models usually used to discuss this
problem are restricted in certain ways. It is assumed that the
bath stays in equilibrium during the times of interest, so that
the internal relaxation times in the bath must be short compared
to the timescale relevant to the dynamics of the quantum system,
and energy given by the quantum system to the bath rapidly moves
away, redistributing itself amongst bath modes.

The problem with such models is that at low $T$ the basic
assumption of short internal relaxation times breaks down. This is
of course well known and has been studied theoretically in, eg.,
spin glasses \cite{hertz,mezard}, dipolar glasses \cite{burin},
some models of low-$T$ phase nucleation \cite{nucleation}, and in
the various relaxation bottlenecks existing in magnetic systems
\cite{abraBl} (to name only a few examples). However the problem
is really generic to low-$T$ physics (a feature well-known to
low-$T$ experimentalists \cite{pobell}, because it is the main
obstacle to cooling to very low $T$). In fact, in almost all
systems apart from pure liquid $^3He$ and $^4He$, the thermal bath
has most of its low $T$ energy and entropy locked up in localised
excitations. The relaxation of energy and entropy in and out of
these modes ranges from $\mu$secs to centuries. From a theoretical
standpoint these facts are not surprising- they often arise in
systems of local modes when couplings and fields are random
(particularly when there is frustration of some kind in the
interactions).

What this means here is that we should reconsider the whole
problem of the quantum-classical crossover, using models which
have these localised modes built in to them from the start.
Consider, eg., the standard 2-well system, coupled now to both
oscillator and spin baths. The oscillators represent delocalised
modes, which at low $T$ are few in number but can move energy
around quickly, and the spin bath represents the localised modes,
which contain all the energy and entropy. A toy model for such a
system is
\begin{eqnarray}
{\cal H} &=& {\cal H}_o(P,Q) + {\cal H}_o^{osc} + {\cal H}_o^{SB}
+ V_{int}  \nonumber \\
{\cal H}_o(P,Q) &=& P^2/2M + U_o(Q) \nonumber \\
 V_{int} &=& Q [\sum_q c_q x_q   + \sum_k
\omega_k^{\parallel} {\it l}_k \cdot \hat{\sigma}_k ]
 \label{Hcross}
\end{eqnarray}
in which the oscillator and spin baths are as before (see eqtn.
(\ref{Hbath})), and the couplings are simple diagonal ones to the
coordinate of the particle. One can have more complicated
couplings, and in general one should also add counterterms
\cite{cal83} to ${\cal H}$ to renormalise the 2-well tunneling
potential back to $U(Q)$.

Already in this toy model one begins to see how the
quantum-classical crossover will work. At low $T$ the dynamics of
the particle will be governed by its coupling to the spin bath, in
the way previously described.  Raising the bath temperature has
little effect on the spin bath dynamics unless the $\{
\omega_k^{\parallel} \}$ are very large- they are already at high
$T$ compared to their basic energy scales $\{
\omega_k^{\parallel}, \omega_k^{\perp} \}$. However eventually the
effect of thermal transitions of the oscillators to higher levels
of the 2-well system begins to take effect. In the absence of the
spin bath one sees a crossover to activated behaviour around a
temperature $T_o \sim \omega_o /2 \pi$, where $\omega_o$ is the
(renormalised) small oscillation frequency of the system in one or
other of the wells \cite{weiss99}. The width $\Delta T_o$ of the
crossover depends on the details of the potential, the bath
coupling, etc., but it will not be less than $\sim O(T_o/n)$,
where $n$ is the number of levels below the barrier.

However the spin bath introduces new timescales in the problem,
viz., the $\tilde{T}_1$ and $\tilde{T}_2$ introduced in the last
section. Now these timescales, depending as they do on the
dynamics of the central system itself, will decrease rapidly as we
raise the temperature- the more rapid fluctuations of the central
system, caused by coupling to thermally excited oscillators,
stimulate more rapid transitions in the spin bath. This indirect
effect of the oscillator bath on the spin bath dynamics, acting
through the central system, is of course well known in NMR. In any
case, we see the spin bath can compete with the oscillator bath
over a rather wide range of temperatures in controlling the
dynamics of the central system. This tells us that we may expect a
much wider crossover between quantum and classical behaviour than
occurs when one only deals with an oscillator bath environment (or
only a spin bath environment).

Rather than give a general study of this crossover here, which is
rather lengthy, we now present instead some relevant results for
magnetic molecular spin relaxation, for which there also exist
fairly detailed experimental results. These results actually
capture some of the more general features of the problem.

\section{4a: Quantum-Classical Crossover for Magnetic Molecules}
\label{sec:4a}

It has been understood for many years that the thermally activated
spin dynamics of insulating magnetic ions and nanomagnets
(including large spin magnetic molecules) is driven by coupling to
the phonon bath. Experimental investigations in the last 10 years
of the tunneling relaxation dynamics of various magnetic molecules
has led to many theoretical attempts to understand the temperature
dependence of this behaviour in terms of the spin-phonon coupling
to single tunneling spins \cite{pol94,garCh97,luis,fort98,loss}.
Such calculations give a typical crossover from some
straightforward low-$T$ tunneling to higher-$T$ activation- the
details tend to be be messy because of the presence of many levels
and different spin-phonon coupling. These calculations must
clearly apply at some sufficiently high temperature that neither
nuclear spins nor intermolecular dipolar interactions are
relevant. However they are not directly applicable to the
experiments in the crossover regime, nor in the quantum regime.
This is clear both on theoretical and experimental grounds, as
follows:

From a theoretical point of view the interplay between nuclear and
phonon couplings on a single nanomagnet cannot be ignored,
especially given that the nuclear dynamics is fast. In experiments
the intermolecular dipole interactions couple the relaxation of
different molecules, so that they can only relax independently
when $k_BT \gg V_D$, where $V_D$ is the strength of these dipolar
interactions. Thus until we reach this rather high temperature
(which in most experiments is well above the temperature $T_o$
defined earlier), both intermolecular dipole and hyperfine
coupling to the nuclear spins must be included on an equal footing
with the spin-phonon interactions.

From the experimental point of view the need for this is obvious.
Even well above $T_o$ the relaxation is non-exponential in time
(as in the quantum regime, but now with $T$-dependent
characteristics), showing the molecules do not relax
independently. Moreover, the 'hole-digging' phenomenon in the
distribution of internal fields, caused by nuclear spins, also
survives at temperatures well above $T_o$ (although with a
time-development that becomes rapidly $T$-dependent),
demonstrating that the nuclear spins are still partially
controlling the tunneling dynamics.

To begin analysis of this problem we note that because the nuclear
spin bath and the phonon bath do not interact with each other
directly (but only via the central molecular spin), we can treat
their relaxation rates as independent. We assume that each
nanomagnet has a spin Hamiltonian of easy axis form $H_o({\bf S})
=H_o^{z}(\hat{S}_z) + V_o^{\perp}(\hat{S}_{\pm})$, so that the
transverse term $V_o^{\perp}$ causes tunneling between the
eigenstates $|m\rangle$ of the longitudinal part (defined by
$H_o^z |m\rangle = E^{(0)}_m|m\rangle$). Let us also assume that
the applied longitudinal field is small, so that levels
$|m\rangle$ and $|-m\rangle$ are near resonance (and level
$|m\rangle$ is not near resonance with any other levels). Now
suppose we start with an ensemble of such nanomagnets, all of them
having $m=S$, ie., so the system is completely polarised. If we
ignore any coherence effects (ie., assume incoherent relaxation),
and also ignore intermolecular interactions, then the kinetic
equation for the system reduces to
\begin{eqnarray}
\dot{P}^{(1)}_m(\xi,{\bf r},t) = &-& \sum_{m'}[\Gamma_{m'm}(\xi,T)
P^{(1)}_m(\xi,{\bf r},t) \nonumber \\ &-& \Gamma_{mm'}(\xi,T)
P^{(1)}_{m'}(\xi, {\bf r},t)]
\label{2.11}
\end{eqnarray}
($\Gamma_{-m,m}=\Gamma_{m,-m} \equiv \Gamma_m$) where
$P^{(1)}_m(\xi,{\bf r},t)$ is the 1-molecule probability
distribution, describing the probability to find a molecule at
position ${\bf r}$, in state $|m\rangle$, in a bias field $\xi$,
at time $t$; and $\Gamma_{mm'}(\xi,T)$ is the rate at which
nanomagnets in a local field $\xi$ make transitions from state
$|m'\rangle$ to $|m\rangle$, under the influence of both phonons
(at a temperature $T$) and nuclear spins. The assumption of
non-interacting phonon and nuclear baths implies we can write:
\begin{equation}
\Gamma_{mm'}(\xi,T) = \Gamma^N_{mm'}(\xi,T) +
\Gamma^{\phi}_{mm'}(\xi,T)
 \label{rate}
\end{equation}
with individually defined nuclear spin- and phonon-mediated
relaxation rates. The system can move up or down levels on the
same side of the barrier by emission of phonons. Of particular
interest here are the inelastic tunneling rates out of level
$|m\rangle$ in a bias field, to the other side of the barrier. The
phonon-mediated process of this kind has the form \cite{KM}:
\begin{equation}
\Gamma^{\phi}_m (\xi) \approx { \Delta^2_m \; W_m(T) \over
\Delta^2_m + \xi^2_m + \hbar^2 W^2_m(T)}
 \label{EPR1}
\end{equation}
where $W_m(T)$ is that part of the linewidth of the $m$-th level
caused by phonon-mediated intra-well processes, $\xi_m = g \mu_B m
H^z$ is the bias energy between levels $ |\pm m\rangle$ (with the
field $H_z$ the sum of internal and external fields), and
$\Delta_m$ is the tunneling matrix element for tunneling between
these same levels.

On the other hand for the nuclear spin bath-mediated rate we will
use here the temperature-{\it independent} form:
\begin{equation}
\Gamma^N_m(\xi) \sim {2 \Delta^2_m G^{(m)}_N \over \pi^{1/2} \hbar
E^{(m)}_o} e^{- |\xi_m| / E^{(m)}_o}
 \label{ENR}
\end{equation}
where $E^{(m)}_o$ is a generalisation of the quantity $E_o$ which
plays a role in nuclear spin-mediated tunneling in the quantum
regime- roughly one has $E^{(m)}_o \sim (m/S) E_o$ (cf., eqtn.
(\ref{Eo})). This quantity measures the range over which the
nuclear spin bias is being swept, either by internal spin
diffusion or by transitions caused by the nanomagnetic dynamics
(the mechanism described in the last section). The factor
$G^{(m)}_N = \exp \{-\Delta^2_m/2 (E^{(m)}_o)^2 \}$ follows from
the Gaussian spread of the nuclear multiplet, and simply says that
$\Gamma^N_m(\xi)$ vanishes when $\Delta_m$ becomes large in
comparison with $E^{(m)}_o$.

The rationale for using (\ref{ENR}) is that we are interested here
in time scales long compared to the $\tilde{T}_1$ and
$\tilde{T}_2$ discussed in the last section. In this case we
expect that the system is able to cover the whole range of states
in the nuclear spin manifold surrounding each level, and so we can
simply weight these according to their number, ie., according to a
density of states. Clearly this approximation breaks down if we
are interested in shorter time scales- we will not consider this
problem here. We can in fact go further- since at time scales
longer than all relaxation times, the nuclear bias will fluctuate
across the whole range of nuclear states, of width $E_0^{(m)}$, we
can incorporate this into the phonon rate as an average- for
example, when $E^{(m)}_0 >>  \max \{ \Delta_m, \hbar W_m \}$, the
phonon rate becomes
\begin{equation}
\Gamma^{\phi}_m(\xi,T) \approx {\Delta^2_m W_m(T)  \over E^{(m)}_0
\sqrt {\Delta^2_m + \hbar^2 W^2_m(T)} } \sqrt {\pi \over 2}
e^{-\xi^2_m / 2 (E^{(m)}_0)^2}.
 \label{EPR2}
\end{equation}
Actually this result turns out to be valid even when the spin bath
dynamics is slow compared to the timescale we are interested in-
but with one simple modification. The point is that even if a
given molecule in a dipolar bias field $\xi_D$, coming from the
other molecules, cannot find resonance, because the spin bath
brings it too slowly to resonance, nevertheless some fraction $x$
of the molecules in a field $\xi$ will find themselves in a
compensating nuclear bias field near $-\xi_D$, near enough so that
the molecule is quickly brought to resonance.

We see that in this very simple approximation the spin and
oscillator bath-mediated relaxation processes already influence
each other strongly, albeit in a rather trivial way. Because
(\ref{EPR1}) reduces the contribution of the higher levels to the
relaxation, which in the usual theory of the crossover take over
very quickly from the lowest levels as one goes through $T = T_o$,
the net effect is to broaden the crossover.

To illustrate this it is useful to show results for a particular
system. We again choose the $Fe$-8 molecule, already discussed
above in the context of low-$T$ decoherence. In Fig. 6 we show the
contributions from the different levels to the relaxation as a
function of $T$ for the $Fe$-8 system. To give results comparable
with experiment we have generalised the kinetic equation
(\ref{2.11}) to include dipolar interactions- this development is
a rather messy but fairly straightforward adaptation of the method
used in ref. \cite{PS98}. The main effect of adding these
interactions is however not to change the width of the crossover,
but rather to change the time dependence of the relaxation- this
it is fairly complex, even in the quantum regime \cite{PS98}, and
not relevant to the present study.

\begin{figure}[ht]
\centering
\vspace{-1.5cm}
\includegraphics[scale=0.35]{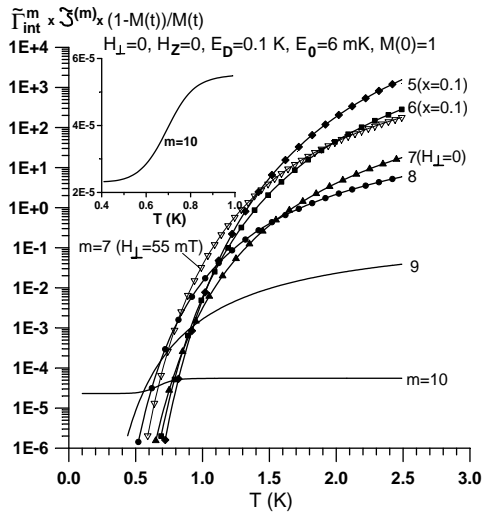}
\vspace{-1.6cm}
\caption{The relaxation rates as a function of temperature
for relevant values of $S_z=m$ in the case of the $Fe_8$ system.
The two curves for $m=7$ correspond to different transverse
fields, viz., (i) $H_{\perp}=0$ (filled triangles); and (ii)
$H_{\perp}=55 \; mT$ (open flipped triangles).
The inset enlarges the $m=10$ curve around $T \gtrsim T_o$, where
$T_o$ is the temperature at which phonon transitions would
normally cause the tunneling dynamics to rapidly crossover to
activated behaviour.}
\label{fig:fig6}
\end{figure}

The most important point to be noted from Fig. 6 is that the
crossover is now very wide- it extends from $T \sim 0.4~K$ up to
$T \sim 1.7~K$, above which temperature the contributions from
level $m=5$ begin to dominate. It will be surprising to those
familiar with the standard theory of the quantum-classical
crossover that it is an intermediate level that dominates. The
crossover is so wide for 2 reasons, viz.,

(i) The reason noted above, ie., the spreading of the levels by
the nuclear spin hyperfine coupling (and in the case of this
calculation, also the dipolar fields, so that now the spread is
even greater than $E_o^{(m)}$). This emphasizes the role of the
lower levels, more than would otherwise be the case; and

(ii) We see already that in the basic phonon transition rate in
(\ref{EPR1}) there is a saturation in the rate for the high levels
(having small $m$ and large $\Delta_m$). This is because the
typical bias $\xi_m$ in this formula, and in the more general
formula (\ref{EPR2}), will now be either a dipolar or nuclear
hyperfine bias, which is much larger than the phonon linewidth
$W_m$. Once $\Delta_m$ exceeds this bias, the presence of
$\Delta_m$ in the denominator stops the rapid increase of the
rate- this happens before one reaches the very highest levels, and
is the basic reason why intermediate levels dominate the
relaxation over a wide temperature range. In the case of $Fe$-8,
this happens for $m=5$, but clearly could happen for some
intermediate level in a different system. In most of the earlier
papers on phonon-mediated tunneling relaxation of nanomagnets, the
factor of $\Delta_m^2$ in the denominator was not included- this
led to a quite different picture of the crossover.

Actually experiments in these systems do show a very wide
crossover. The detailed comparison between theory and experiment
is rather interesting, since one may analyse both the relaxation
as a function of time (and how the form changes with $T$, along
with rate) and also the $T$-dependence of the hole-digging
dynamics \cite{tupCross}.

\section{Summary}
\label{sec:5}

This paper has not attempted a complete study- instead we have
tried to make some general points about 2 kinds of crossover,
uncluttered by too much detail for particular systems. To
illustrate the general remarks it has been nevertheless useful to
give results for the $Fe$-8 molecule. It goes without saying that
the detailed calculations for this system (and others, such as
SQUIDs) are rather lengthy, and including them would have obscured
the points we wishes to make.

The 2 main points are that

(i) The presence of the spin bath is very bad for decoherence when
the basic energy scale $\Delta_o$ of a qubit is small. This means
that the spin-boson model is not applicable at all in this regime-
one must use a 'central spin' model \cite{PS00}. However if we
increase $\Delta_o$, the spin bath decoherence effects fall off
extremely rapidly, and eventually become negligible. At this point
the spin-boson model becomes applicable, and decoherence begins to
increase again as one further increases $\Delta_o$, a well-known
feature of the model \cite{weiss99}.

(ii) In the standard quantum-classical crossover that occurs as
one raises the bath temperature for a tunneling system, the
presence of the spin bath is again very important. The main effect
it has is first to completely change behaviour in the quantum
regime, and then to enormously broaden the usual rather sharp
crossover that exists when one only deals with an oscillator bath.


It is quite clear that these results are only the beginning of a
proper study of the way in which spin and oscillator baths work
together. The remarkable edifice of theoretical work that has been
constructed around the spin-boson and related models
\cite{weiss99} should be a very nice model for what interesting
paths remain to be explored in this area.

\vspace{4mm}

We thank J. Angles d'Auriac and J. de Jongh for hospitality while
this article was being written, and the CIAR and NSERC in Canada,
and grant number NS-1767.2003.2 in Russia, for support.

\end{document}